\documentclass[]{interact}

\usepackage{epstopdf}
\usepackage[caption=false]{subfig}
\usepackage{hyperref}
\usepackage[numbers,sort&compress]{natbib}
\bibpunct[, ]{[}{]}{,}{n}{,}{,}
\renewcommand\bibfont{\fontsize{10}{12}\selectfont}

\theoremstyle{plain}

\theoremstyle{definition}

\theoremstyle{remark}

\begin{document}

\articletype{Original Research Article}

\title{Diagnostics for  categorical response models based on quantile residuals and distance measures}

\author{
\name{Patr\'{i}cia Peres Araripe \textsuperscript{a}, Idemauro Antonio Rodrigues de Lara\textsuperscript{a}\thanks{Corresponding author: idemauro@usp.br}, Gabriel Rodrigues Palma\textsuperscript{b},  Niamh Cahill\textsuperscript{b} and Rafael de Andrade Moral\textsuperscript{b}}
\affil{\textsuperscript{a}``Luiz de Queiroz'' College,  Postgraduate Program in  Statistics and Agronomic Experimentation, University of S\~{a}o Paulo,  Brazil; \textsuperscript{b}Department of Mathematics and Statistics, Maynooth University, Ireland}
}

\maketitle

\begin{abstract}
Polytomous categorical data, in a nominal or ordinal scale, are frequent in many studies in different areas of knowledge. Depending on experimental design, these data can be obtained with an individual or grouped structure. In both structures, the multinomial distribution may be suitable to model the response variable and, in general, the generalized logit model is commonly used to relate the covariates' potential effects on the response variable. After fitting a multi-categorical model, one of the challenges is the definition of an appropriate residual and choosing diagnostic techniques to assess goodness-of-fit, as well as validate inferences based on the model. Since the polytomous variable is multivariate, raw, Pearson, or deviance residuals are vectors and their asymptotic distribution is generally unknown, which leads to potential difficulties in graphical visualization and interpretation. Therefore, the definition of appropriate residuals, as well as the choice of the correct analysis in diagnostic tools is very important, especially for nominal categorical data, where a restriction of methods is observed. This paper proposes the use of randomized quantile residuals associated with individual and grouped nominal data, as well as Euclidean and Mahalanobis distance measures associated with grouped data only, as an alternative method to reduce the dimension of the residuals and to study outliers. To show the effectiveness of the proposed methods, we developed simulation studies with individual and grouped categorical data structures associated with generalized logit models. Parameter estimation was carried out by maximum likelihood and half-normal plots with simulation envelopes were used to assess model performance using residuals and distance metrics. These studies demonstrated a good performance of the quantile residuals and, also, the distance measurements allowed a better interpretation of the graphical techniques. We illustrate the proposed procedures with two applications to real data, for which the employed techniques validated the model choice.

\end{abstract}

\begin{keywords}
Generalized logit model; maximum likelihood; model selection; half-normal plot; normality.
\end{keywords}

\section{Introduction}

Nominal polytomous variables are defined by a finite set of categories (more than two), being of interest in experimental design in many disciplines, especially in biological and agricultural sciences. The subjects can be an individual (a plant, an insect, or an animal) or groups of individuals (a stall with animals, a cage with insects, or a plant with its branches), in which the categorical responses are observed. Therefore, the categorical data structure can be individual or grouped, depending on experimental design and goals of the study.
Independent of structure, in general, the multinomial distribution (or an extension of it) is assumed to model the response variable, and the generalized logit model is used to describe the relationship between the polytomous response and covariates \cite{agresti}. 

Additionally, the assumptions of the fitted model must be verified to validate statistical inference. In this process, residual analysis is fundamental. The first step involves an appropriate definition for the residuals and, after that, formal (hypothesis tests) and informal (exploratory plots) techniques can be used to assess goodness-of-fit and model assumptions. According to \cite{feng2020}, residual analyses are essential to identify discrepancies between the model and the data, detecting outliers and influential points. For example, the deviance and Pearson statistics are quantitative measures widely used to test the goodness-of-fit of generalized linear models (GLMs), however they can only be applied to multinomial data in the grouped structure and are not reliable for small sample sizes \cite{tutz2011regression}. In fact, residual analysis is still a challenge for the multinomial case. Since the polytomous response is multivariate, the ordinary residual, defined by the difference between the observed response and the estimated probabilities, is a vector for each individual, with a dimension defined by the number of categories. In addition, these residuals have an unknown asymptotic distribution, making them difficult to interpret in diagnostic plots \cite{reiter2005}. It is important, therefore, to find or adapt diagnostic techniques to overcome these limitations.

A few alternatives have been proposed: the first one is to reduce the number of categories (grouping into two) and to do residual analysis by means of logistic regression, whose techniques are well consolidated (e.g. \citealp{pregibon1981logistic}, \citealp{landwehr1984graphical}, \citealp{hossain2003application}). However, grouping categories leads to the loss of information. Another would be to fit the generalized logit models (for pairs of variables) separately, and to define residuals for each sub-model and apply diagnostic tools, as proposed by \cite{silva2003metodos} with three categories.  However, the maximum likelihood estimates from the separate fitted models differ from those obtained in the simultaneous fit, and their standard errors tend to be larger \cite{agresti}.

For ordinal responses, \cite{cheng2021surrogate} defined a continuous residual vector for the individual structure, with three categories, based on the methodology of \cite{liu2018residuals}. Also, they presented the deviance and Pearson residual vectors and plots of residuals versus covariates. However, this technique is not suitable to the nominal case. For nominal data with grouped structure, \cite{seber2000residuals} defined a vector of residuals based on the projected residuals presented by \cite{cook1985residualsnew}, and the Pearson residuals were presented by \cite{gupta2008residuals} to detect influential points. However, these methodologies require theoretical development and are not implemented in statistical software.

A residual defined for a broad class of models that can be easily implemented is the randomized quantile residual \cite{dunn1996}. For discrete data, these residuals are an extension of the quantile residuals for continuous data and they follow approximately a normal distribution if the estimated parameters are consistent, but it is important to investigate their properties in small sample sizes \cite{pereira2019quantile}. \cite{feng2020}. Therefore, the quantile residuals are an alternative for multinomial case associated to generalized logit models, but there is a lack of investigation of their performance. In addition to the adoption of quantile residuals, another alternative is to use distance metrics, such as Euclidean and Mahalanobis distances, to reduce the dimension of the ordinary residuals for diagnostic analyses. These metrics are widespread in the literature on multivariate analysis, when calculating how far two individuals are in the original variable space (e.g. by using principal components and cluster analysis) \cite{johnson2007applied}. In the context of diagnostics, these distances have already been used to detect outliers in linear regression (\citealp{hadi2009detection} and \citealp{ghorbani2019mahalanobis}). However, there are no records of their use in models for nominal data.

Here, we propose the adoption of quantile residuals and the use of multivariate distance metrics. Our objectives are: (i) to assess the normality of randomized quantile residuals for nominal categorical models; and (ii) to reduce the dimension of ordinary residuals associated with nominal data using Euclidean and Mahalanobis distances for grouped structures. We review models and residuals for nominal polytomous data in Section \ref{rev}. Then, we present the randomized quantile residual  and the distance metrics in Sections \ref{qr} and \ref{dist}, respectively. The framework based on randomized quantile residuals and distances for nominal responses, which are the contributions of this work, are presented in Section \ref{met}. We present results of simulation studies in Section \ref{simul}, and illustrate with two applications from the literature in Section \ref{aplic}. Finally, we present concluding remarks in Section \ref{final}. 

\section{Models and residuals for nominal polytomous data} \label{rev}

Statistical models for polytomous data (nominal or ordinal) are based on the multinomial probability distribution \cite{agresti2007introduction}. The definition of the linear predictor structure is essential when defining the model, and influences the construction of residuals, as well as the diagnostic techniques.

\subsection{Nominal data structures}

It is important to distinguish between individual and grouped data structures. To establish the notation consider a sample of subjects, $i= 1, 2, \ldots, n$ and the set of $J$ categories $A=\{1,2, \ldots, J\}$. In the individual case, each subject is a single individual, which is classified in some category of set $A$. Then, the random vector referring to the individual $i$ is given by ${\mathbf{Y}}_i = (Y_{i1},\ldots,Y_{iJ})'$, where $Y_{ij}=1$ if the response of individual $i$ is in category $j$, $j=1,2,\ldots,J$, and $Y_{ij}=0$ otherwise, with $\sum\limits_{j = 1}^J {{Y_{ij}} = 1}$. For the grouped case, each subject is composed of a group of $m_i$ individuals. Then, the random variable $Y_{ij}$ represents the number of times category $j$ was observed in $m_i$ individuals, with $\sum\limits_{j = 1}^J {{Y_{ij}} = m_i}$. 
In both cases, we have a multinomial trial, that is, it is assumed that the random vector
 ${\mathbf{Y}}_i$ follows a multinomial distribution, ${\bf{Y}}_i \sim \mbox{Multi}(m_i,{\mbox{\boldmath{$\pi$}}}_i)$, with parameters $ m_i $ and ${\mbox{\boldmath{$\pi$}}}_i=(\pi_{i1},\ldots,\pi_{iJ})'$, restricted to $\sum\limits_{j = 1}^J {{{\pi _{ij}}} = 1}$.

\subsection{Generalized logit model} \label{glm}

The multinomial distribution belongs to the canonical multi-parametric exponential family, with a vector of canonical parameters $\mbox{\boldmath{$\theta$}} = {\left[ {\log \left( {\frac{{{\pi _1}}}{{{\pi _J}}}} \right), \ldots ,\log \left( {\frac{{{\pi _{J- 1}}}}{{{\pi _J}}}} \right)} \right]'}$, where ${\theta _j} = \log \left( {\frac{{{\pi _j}}}{{{\pi _J}}}} \right)$, $j=1,\ldots,J-1$, use canonical link functions. Considering a random sample of subjects of dimension $n$,  $i = 1,2,\ldots, n$, the generalized logit model is defined as
\begin{equation}
 {\rm{logit}}\left[ {{\pi _{ij}}{({\bf{x}}_i)}} \right] = \log \left[ {\frac{{{\pi _{ij}}}{({\bf{x}}_i)}}{{{\pi _{iJ}}{({\bf{x}}_i)}}}} \right] = {\alpha _{j}} + \sum\limits_{k = 1}^p {{\beta _{jk}}{x_{ik}}}= {\alpha _{j}} +{\mbox{\boldmath{$\beta$}}}{_j'}{\bf{x}}_i,\,\,\,\,j = 1, \ldots ,J - 1,
 \label{eq5}
\end{equation}
where $J$ is the number of categories, ${\pi _j}{({\bf{x}}_i)}$ is the probability of an individual response  in the $j$-th category, ${\bf{x}}_i = ({x_{i1}},{x_{i2}},\ldots ,{x_{ip}})'$ is the vector of covariates, ${\mbox{\boldmath{$\beta$}}}{_j}= ({\beta_{j1}},{\beta_{j2}},\ldots ,{\beta_{jp}})'$ represents the vector of parameters, and ${\alpha _{j}}$ is the intercept. According to \cite{agresti}, the covariates can be quantitative, factors (using dummy variables) or both. Model \ref{eq5} compares each category with one chosen as a reference, generally the first or last category, but this choice can be arbitrary \cite{tang2012} \cite{bilder}. Also, from equation~(\ref{eq5}), we have:
\begin{equation} \label{prob1}
{\pi_{ij}}{({\bf{x}}_i)} = \frac{{\exp \left( {{\alpha_{j}}  +{\mbox{\boldmath{$\beta$}}}{_j'}{\bf{x}}_i} \right)}}{{1 + \sum\limits_{j = 1}^{J - 1} {\exp \left( {{\alpha_{j}}  +{\mbox{\boldmath{$\beta$}}}{_j'}{\bf{x}}_i} \right)} }},\,\,\,\,j = 1, \ldots ,J - 1,
\end{equation}
and the probability for the reference category: 
\begin{equation} \label{prob2}
	{\pi_{iJ}}{({\bf{x}}_i)} =  1 - \left[ {{\pi _{i1}}({\bf{x}}_i) +  \cdots  + {\pi _{i(J - 1)}}({\bf{x}}_i)} \right]=\frac{1}{{1 + \sum\limits_{j = 1}^{J - 1} {\exp \left( {{\alpha _{j}}  +{\mbox{\boldmath{$\beta$}}}{_j'}{\bf{x}}_i} \right)} }}.
\end{equation} 
The parameter estimation process is done by maximum likelihood, which consists of maximizing ${\pi_{ij}}{({\bf{x}}_i)}$ to simultaneously satisfy the $J-1$ equations that specify the model. 

We present here a brief summary of the process, distinguishing the individual and grouped structures. 
First, consider the data with individual structure with the observed vector ${{\bf{y}}_i} = ({y_{i1}}, \ldots ,{y_{iJ}})$ satisfying $\sum\limits_{j = 1}^J {{y_{ij}} = 1} $, with mean $\mbox{E}(Y_{ij}|{\mathbf{x}}_i)=\pi_{ij}({\mathbf{x}}_i)$, $j = 1, 2,\ldots, J$. Then, the log-likelihood function is given by 
\[l = \log \prod\limits_{i = 1}^n {\left\{ {\prod\limits_{j = 1}^J {{{\left[ {{\pi _{ij}}\left( {{{\bf{x}}_i}} \right)} \right]}^{{y_{ij}}}}} } \right\}}  = \log \prod\limits_{i = 1}^n {\left\{ {\prod\limits_{j = 1}^{J - 1} {{{\left[ {{\pi _{ij}}\left( {{{\bf{x}}_i}} \right)} \right]}^{{y_{ij}}}}{{\left[ {{\pi _{iJ}}\left( {{{\bf{x}}_i}} \right)} \right]}^{{y_{iJ}}}}} } \right\}}. \]
Using~(\ref{prob1}) and~(\ref{prob2}) we have
\[\begin{array}{l} 

l = \sum\limits_{i = 1}^n {\left\{ {\sum\limits_{j = 1}^{J - 1} {{y_{ij}}\left( {{\alpha _{j}} + {\mbox{\boldmath{$\beta$}}}_{j}'{\bf{x}}_i} \right)}  - \log \left[ {1 + \sum\limits_{j = 1}^{J - 1} {\exp ({\alpha _{j}} + {\mbox{\boldmath{$\beta$}}}_{j}'{{\bf{x}}_i})}}  \right]}\right\}}. 
\end{array}\]

Now, considering the grouped data where the observed vector ${{\bf{y}}_i} = ({y_{i1}}, \ldots ,{y_{iJ}})$ satisfies $\sum\limits_{j = 1}^J {{y_{ij}} = m_i}$, with mean $\mbox{E}(Y_{ij}|\mathbf{x}_i)=m_i\pi_{ij}({\mathbf{x}}_i)$, $j=1,\ldots,J$, the 
log-likelihood is given by
\[\begin{array}{l}
l^* = \sum\limits_{i = 1}^n {\left\{ {\sum\limits_{j = 1}^J {{y_{ij}}\log \left[ {{\pi_{ij}} \left( {{{\bf{x}}_i}} \right)} \right] + \log \left[ {\frac{{{m_i}!}}{{{y_{i1}}! \ldots {y_{iJ}}!}}} \right]} } \right\}} \\
\,\,\,\,\,\, = \sum\limits_{i = 1}^n {\left\{ {\sum\limits_{j = 1}^{J - 1} {{y_{ij}}\log \left[ {{\pi _{ij}}\left( {{{\bf{x}}_i}} \right)} \right] + {y_{iJ}}\log \left[ {{\pi _{iJ}}\left( {{{\bf{x}}_i}} \right)} \right] + \log \left[ {\frac{{{m_i}!}}{{{y_{i1}}! \ldots {y_{iJ}}!}}} \right]} } \right\}} .
\end{array}\]
and, similarly to the individual process, by successive substitutions one finds:
\[\begin{array}{l}
l^* = \sum\limits_{i = 1}^n {\left\{ {\sum\limits_{j = 1}^{J - 1} {{y_{ij}}({\alpha _{j}} +{\mbox{\boldmath{$\beta$}}}_{j}'{\bf{x}}_i)  - {m_i}\log \left[ {1 + \sum\limits_{j = 1}^{J - 1} {\exp ({\alpha _j} +{\mbox{\boldmath{$\beta$}}}_{j}'{{\bf{x}}_i}} )} \right] + \log \left[ {\frac{{{m_i}!}}{{{y_{i1}}! \ldots {y_{iJ}}!}}} \right]} } \right\}}. 
\end{array}\]
An iterative method such as Newton-Raphson can be used to maximize $l$ and $l^*$ to obtain the maximum likelihood estimates \cite{tutz2011regression}. More details can be found in
\cite{agresti2007introduction}.

\subsection{Residuals associated with models for nominal categorical data}
An important step in model diagnostic checking is residuals analysis, used to validate model assumptions and detect outliers or influential points \cite{singer2007res}. The definition of the residuals as well as the analytical techniques are essential tools that contribute to this.

\subsubsection{Residuals for individual data}
The ordinary residuals measure the deviations between the observed values and the predicted probabilities. For model~(\ref{eq5}) they are vectors of dimension $J\times 1$ per individual, $i=1,2,\ldots,n$, given by \cite{reiter2005}
\[
{\hat{\bf{r}}}_i ={{\bf{y}}_i}-\hat{\mbox{\boldmath{$\pi$}}}_i=\left( {{y_{i1}} - {{\hat \pi }_{i1}},{y_{i2}} - {{\hat \pi }_{i2}}, \ldots ,{y_{iJ}} - {{\hat \pi }_{iJ}}} \right)',\]
where ${\bf{y}}_i=(y_{i1},y_{i2},\ldots,y_{iJ})'$ is the vector of observations with $y_{ij}=1$ if the individual response  $i$ belongs to category $j$ and $y_{ij}=0$, otherwise, and  ${\mbox{\boldmath{$\pi$}}}_i=({{\hat \pi }_{i1}},{{\hat \pi }_{i2}},\ldots,{{\hat \pi }_{iJ}})'$ is the vector of predicted probabilities. These residuals do not follow a multivariate normal distribution, and when used in diagnostic plots, they may not be informative, since visual interpretation is not straightforward.

The Pearson and deviance residuals for model~(\ref{eq5}) are given, respectively, by the vectors ${r_i^P} = {\left[ {r_{i1}^P,r_{i2}^P, \ldots, r_{iJ}^P} \right]^\prime }$ and ${r_i^D} = {\left[ {r_{i1}^D,r_{i2}^D, \ldots, r_{iJ}^D} \right]^\prime }$, whose elements are obtained by  \cite{cheng2021surrogate}
\[r_{ij}^P = \frac{{\left( {{y_{ij}} - {{\hat \pi }_{ij}}} \right)}}{{\sqrt {{{\hat \pi }_{ij}}(1 - {{\hat \pi }_{ij}})} }}\]
and
\[{r_{ij}^D} = \mbox{sign}\left( {{y_{ij}} - {{\hat \pi }_{ij}}} \right)\sqrt {2\left[ {\left( {{y_{ij}-1}} \right)\log \left( {1 - {{\hat \pi }_{ij}}} \right) - {y_{ij}}\log \left( {{{\hat \pi }_{ij}}} \right)} \right]}, \]
where $j=1,2,\ldots,J$. These definitions are  extensions of the residuals used in logistic regression.
Specifically for variables on the ordinal scale,
 \cite{cheng2021surrogate} proposed the surrogate residuals, that are based on the methodology presented by \cite{liu2018residuals}.
 As the scope of this work is centered on the nominal measurement scale, we leave it to the interested readers to consult \cite{cheng2021surrogate} for more details.

\subsubsection{Residuals for grouped data} \label{resgrup}
The $J$-dimensional ordinary residuals vector for model~(\ref{eq5}) per subject $i$, $i=1,2,\ldots,n$, each with $m_{i}$ individuals, according to  \cite{tutz2011regression} is defined by
\begin{eqnarray*}
\hat{\bf{r}}_i &=& \frac{{{\bf{y}}_i}-m_i \times\hat{\mbox{\boldmath{$\pi$}}}_i}{m_i} \\
&=& \frac{1}{m_i}{\left( {{y_{i1}} - m_i{{\hat \pi }_{i1}},{y_{i2}} - m_i {{\hat \pi }_{i2}}, \ldots ,{y_{iJ}} - m_i{{\hat \pi }_{iJ}}} \right)',}
\end{eqnarray*}
where ${\bf{y}}_i=(y_{i1},y_{i2},\ldots,y_{iJ})'$ is the vector of observed counts, such that  $\sum\limits_{j = 1}^J {{y_{ij}} = m_i}$, and  $\hat{\mbox{\boldmath{$\pi$}}}_i=({{\hat \pi }_{i1}},{{\hat \pi }_{i2}},\ldots,{{\hat \pi }_{iJ}})'$ is the vector of predicted probabilities.
The $J$-dimensional vector of Pearson residuals is given by ${r_i^P} = {\left[ {r_{i1}^P,r_{i2}^P, \ldots, r_{iJ}^P} \right]' }$ with elements \citep{tutz2011regression}
 \[r_{ij}^P = \frac{{\left( {{y_{ij}} - m_i {{\hat \pi }_{ij}}} \right)}}{{\sqrt {m_i{{\hat \pi }_{ij}}(1 - {{\hat \pi }_{ij}})} }},\]
where $i=1,2,\ldots,n$ and $j=1,2,\ldots,J$.

\section{Randomized quantile residuals}\label{qr}
The quantile residual was proposed by \cite{dunn1996} for continuous variables. For a continuous response, $y_i$, the quantile residual is defined by
\[r_i^Q = {\Phi ^{ - 1}}\left\{ {F({y_i};{{\hat \theta }_i},\hat \phi) } \right\},\,\,\,i = 1, \ldots ,n,\]
where ${\Phi ^{ - 1}}$ is the inverse of the cumulative distribution function (CDF) of the standard normal distribution, ${F({y_i};{{\hat \theta }_i},\hat \phi)}$ is the CDF associated with the response variable, ${{\hat \theta }_i}$ is the maximum likelihood estimate of parameter $\theta_i$ and the $\hat \phi$ is the estimated dispersion parameter.

If the response $y_i$ is discrete, we introduce randomization through a uniform random variable in the CDF for each individual, obtaining the randomized quantile residual
\[r_i^Q = {\Phi ^{ - 1}}\left\{ {F({u_i}) } \right\},\,\,\,i = 1, \ldots ,n,\]
where $u_i$ represents a uniform random variable between ${a_i} = {\lim _{y \to {y_i}}}F(y;{{\hat \theta }_i},\hat \phi)$ and ${b_i} = F({y_i};{{\hat \theta }_i},\hat \phi)$. Under a well-fitting model, these residuals follow, approximately, a normal distribution.

The quantile residuals have received little attention in the literature as model diagnostic tools until recently. For example, \cite{klar2012} used the standardized quantile residuals in goodness-of-fit tests for generalized linear models with inverse Gaussian and gamma variables. \cite{pereira2019quantile} investigated the performance of the quantile residual for diagnostics of the beta regression model and  \cite{feng2020} used the standardized randomized quantile residuals to examine the goodness-of-fit of models applied to count data. Here, we introduce their use with polytomous data associated with generalized logit models.

\section{Distances} \label{dist}

Consider having $n$ individuals denoted by the random vectors ${\bf{z}}_i = \left( {{z_{i1}},{z_{i2}}, \ldots ,{z_{iq}}} \right)'$, $i=1,2,\ldots,n$.  Each individual is represented by a point in $q$-dimensional space, with each dimension representing a variable \cite{sharma}. Distance metrics can quantify how far two individuals are by a scalar which measures their proximity. The Euclidean and Mahalanobis distances are widely known (see  \cite{zelterman2015} and \cite{kannan2015outlier}) and can be calculated in the original scale of the response variable \cite{de2000mahalanobis}. 
The Euclidean distance between individuals $i$ and $t$ is defined by
\[d_{it}^E = \sqrt {({{\bf{z}}_i} - {{\bf{z}}_t})'({{\bf{z}}_i} - {{\bf{z}}_t})}  = \sqrt {\sum\limits_{k = 1}^q {\left( {{z_{ik}} - {z_{tk}}} \right)^{2}} } ,\]
where $z_{.k}$ is the $k$-th variable, with $k=1,2,\ldots, q$, and $i, t=1,2,\ldots, n$. According to \cite{zelterman2015}, this measure is the most popular to calculate the distance between individuals in $q$-dimensional space.

If the individuals are correlated, the covariance or correlation between them can be considered when calculating the distance \cite{sharma}. In this case, the Mahalanobis distance is useful, and is expressed by
\[d_{it}^M = {({{\bf{z}}_i} - {{\bf{z}}_t})^\prime }{{\bf{C}}^{ - 1}}({{\bf{z}}_i} - {{\bf{z}}_t}),\]
where ${{\bf{C}}^{ - 1}}$ is the inverse of the $q \times q$ variance-covariance matrix. In the case where  ${{\bf{C}}}=\bf{I}$, with $\bf{I}$ representing the identity matrix, the Mahalanobis distance reduces to the Euclidean distance. If ${{\bf{C}}}$ is a diagonal matrix, then it results in the standardized Euclidean distance \cite{johnson2007applied}.
The Euclidean distance yields quicker calculations than the Mahalanobis distance, but considering the covariances between variables can be important \cite{ghorbani2019mahalanobis}. However, \cite{de2000mahalanobis} reported that some issues must be observed when using Mahalanobis distances, such as problems that may lead to singular covariance matrices and the restriction that the sample size that must be greater than the number of variables.

\section{Methods} \label{met}
Here, we describe the methodological procedures associated with residual analysis of generalized logit models fitted to nominal polytomous data. We propose the use of quantile residuals for individual data, and a new methodology to reduce the dimension of ordinary residuals associated with grouped data using distance metrics.

\subsection{Individual data}

For individual data, we obtain the standardized randomized quantile residuals considering the cumulative distribution function (CDF), $F({\bf{y}}_i;{\hat{{\mbox{\boldmath{$\pi$}}}}}_i,\hat{\phi})$, for the response vector ${\mathbf{y}}_i$ given the vector ${\mathbf{x}}_i$, $i=1,2,\ldots,n$. The CDF for the multinomial distribution follows from its relationship with an independent Poisson sum, given a fixed total, i.e., the multinomial CDF is computed as the convolution of $J$ truncated Poisson random variables, as shown by \cite{levin1981} and implemented in R through the \texttt{pmultinom} package (\cite{pmultinom}).


Now let ${\hat{{\mbox{\boldmath{$\pi$}}}}}_i=\left( {{{\hat \pi }_{i1}}({x_i}),{{\hat \pi }_{i2}}({x_i}), \ldots ,{{\hat \pi }_{iJ}}({x_i})} \right)'$ be the vector of estimated probabilities. Consider the probability mass function $f({\bf{y}}_i;{\hat{{\mbox{\boldmath{$\pi$}}}}}_i)$, corresponding the response of individual $i$ in category $j$, $y_{ij} = 1$, and $y_{ij} = 0$ otherwise. Then, the estimated CDF for individual $i$ is
\begin{eqnarray}\label{cdf}
F^*({\bf{y}}_i,u_i;{\hat{{\mbox{\boldmath{$\pi$}}}}}_i)=F(\mathbf{1}-{\bf{y}}_i; {\hat{{\mbox{\boldmath{$\pi$}}}}}_i)+u_i \times f({\bf{y}}_i;{\hat{{\mbox{\boldmath{$\pi$}}}}}_i),
\end{eqnarray}
where $\mathbf{1}$ is a $J\times1$ unit vector, and $u_i$ is a realization of a random variable with uniform distribution, i.e. $U_{i} \sim (0,1)$. The randomized quantile residual for a polytomous response ${\bf{y}}_i$ is given by
\begin{equation} \label{quant_res}
r_i^Q= \Phi^{-1}[F^*({\bf{y}}_i,u_i;{\hat{{\mbox{\boldmath{$\pi$}}}}}_i)],    
\end{equation}
where ${\Phi}^{-1}$ is the quantile function of the standard normal
distribution. We have therefore a scalar value for each $i$, and these residuals are approximately normal under the null hypothesis that the model was correctly specified.

Here, we used a standardized version of the randomized quantile residuals, given by
\begin{equation}\label{quant_res_pad}
r_i^S = \frac{{r_i^Q - {{\bar r}^Q}}}{{{s_{{r^Q}}}}},    
\end{equation}
where ${\bar r^Q}$ and $s_{{r^Q}}$ are the mean and standard deviation of the residuals $r^Q_i$, respectively.

For individual data, distance measurements do not effectively contribute to the analysis of residuals, since regardless of the number of individuals in the sample, the individual structure always leads to $J$ unique distance measurements, under the assumption that the model is correctly specified, i.e, $\mbox{E}({\bf{r}}_i|{\bf{x}}_i)={\bf{0}}$.

\subsection{Grouped data}

For grouped data we can use a similar procedure to construct the randomized quantile residuals~(eq.~\ref{quant_res_pad}),  with the following modification in the estimated CDF  (eq. \ref{cdf}): 
\[ F^*({\bf{y}}_i,u_i;{\hat{{\mbox{\boldmath{$\pi$}}}}}_i)=F(\mathbf{m}-{\bf{y}}_i; {\hat{{\mbox{\boldmath{$\pi$}}}}}_i)+u_i \times f({\bf{y}}_i;{\hat{{\mbox{\boldmath{$\pi$}}}}}_i),\]
where $\mathbf{m}$ represents a $J\times1$ vector for group size, ${\bf{y}}_{i}$ represents the counts vector for each category in the group $i$, in which the sum is $m$.
Additionally, unlike the individual case,  we reduce the dimension of the vector of ordinary residuals using distance metrics, namely the Euclidean and Mahalanobis distances. Under the assumption that the model is specified correctly, we have that $\mbox{E}({\bf{r}}_i|{\bf{x}}_i)={\bf{0}}$, which is a null vector of dimension $J$. We have that the Euclidean and Mahalanobis distances between the residual vector $i$ and the null vector are, respectively, written as
$${d_{i}^{E}} = \sqrt {({\bf{r}}_i - {\bf{0}})'({\bf{r}}_i - {\bf{0}})}= \sqrt{\sum\nolimits_{j = 1}^J {r^2_{ij}} }
$$
and
\[d^M_{i} = ({\bf{r}}_i - {\bf{0}})'{{\bf{C}}^{ - 1}}({\bf{r}}_i - {\bf{0}})= {\bf{r}}_i'{{\bf{C}}^{ - 1}}{\bf{r}}_i,\]
where ${{\bf{C}}}$ is the $J \times J$ covariance matrix of the residuals.

\subsection{Residual analytic tools}

Once the randomized quantile residuals and distance measures are defined, formal (tests) and informal (plots) techniques are employed for diagnostics. Formally, a powerful and widely known test for detecting deviations from normality due to asymmetry or kurtosis (or both) is the Shapiro-Wilk test \cite{shapiro1965}.

Informally, one can first visualize the distribution of residuals through a histogram, comparing its shape with that of the normal distribution. In the plot of residuals versus fitted values, it is possible to observe the existence of variance heterogeneity or the presence of outliers. The expected pattern in this plot is the zero-centered distribution of residuals with constant amplitude \cite{faraway2016}.
 
Additionally, the half-normal plot with a simulated envelope can be used to assess whether the observed data are a plausible realization of the fitted model. The absolute values of a given diagnostic measure (residuals or distances) are compared to the expected order statistics of the half-normal distribution obtained by
\[{\Phi ^{ - 1}}\left[ {\frac{{\left( {i + n - {1 \mathord{\left/
 {\vphantom {1 8}} \right.
 \kern-\nulldelimiterspace} 8}} \right)}}{{2n + {1 \mathord{\left/
 {\vphantom {1 2}} \right.
 \kern-\nulldelimiterspace} 2}}}} \right],\]
where ${\Phi ^{ - 1}}$ is the standard normal quantile function. Here, we follow the steps established by \cite{moral2017half} for the construction of these graphs. Given a well-fitted model, we expect most points to lie within the simulated envelope.

\section{Simulation studies} \label{simul}

We carried out simulation studies to evaluate the performance of the standardized quantile residuals for individual and grouped data, as well as distance measures for grouped data only.

\subsection{Models and scenarios}

We simulated from generalized logit models with 3, 4 and 5 response categories for both data structures (individual and grouped). We used two types of linear predictors: one with an intercept and a single continuous covariate effect (eq.~\ref{model1}), and one also including the effect of a factor with two levels (eq.~\ref{model2}). In the data simulation process, we considered sample sizes of 50, 100 and 200, and for grouped data we used the group dimensions $m\in\{5,10,15\}$. 

For model 1 the response variables were simulated from:
\begin{eqnarray}\label{model1}
\log \left( {\frac{{{\pi _{ij}}}}{{{\pi _{i1}}}}} \right) = {\alpha _{j}} + {\beta _{j}}x_i,\,\,\,\,j=2,\ldots,J,    
\end{eqnarray}
where $x_i$ are realizations of a standard normal random variable, and $J=3,4,5$ according to the number of categories. 
The true parameter values were set as: 
\begin{eqnarray}\nonumber
{\boldsymbol{\theta}}_{(J=3)}&=&(\alpha_{2}, \alpha_{3}, \beta_{2}, \beta_{3})\\\nonumber
&=&(1.5, 3.0, -3.0, -5.0)\\\nonumber
\end{eqnarray}
\vspace{-1.7cm}
\begin{eqnarray}\nonumber
{\boldsymbol{\theta}}_{(J=4)}&=&(\alpha_{2}, \alpha_{3}, \alpha_{4}, \beta_{2}, \beta_{3}, \beta_{4})\\\nonumber
&=&(1.5, 3.0, 2.0, -3.0, -5.0, -4.0)\\\nonumber
\end{eqnarray}
\vspace{-1.7cm}
\begin{eqnarray}\nonumber
{\boldsymbol{\theta}}_{(J=5)}&=&(\alpha_{2}, \alpha_{3}, \alpha_{3}, \alpha_{4}, \beta_{2}, \beta_{3}, \beta_{4}, \beta_{4})\\\nonumber
&=&(1.5, 3.0, 2.0, 4.0, -3.0, -5.0, -4.0, -7.0)\nonumber
\end{eqnarray}



For model 2, the linear predictor was:
\begin{equation}\label{model2}
\log \left( {\frac{{{\pi _{ij}}}}{{{\pi _{i1}}}}} \right) = {\alpha _{j}} +{\beta_{1j}}x_{i1}+{\beta_{2j}}x_{i2},\,\,\,\,j=2, \dots,J,
\end{equation}
where $x_{i1}$ are realizations of a standard normal random variable, $x_{i2}$ is a dummy variable (factor with two levels), and and $J=3,4,5$ according to the number of categories. The true values used were:
\vspace{-0.3cm}
\begin{eqnarray}\nonumber
{\boldsymbol{\theta}}_{(J=3)}&=&(\alpha_{2}, \alpha_{3}, \beta_{12}, \beta_{13}, \beta_{22},\beta_{23})\\\nonumber
&=&(1.5, 3.0, -3.0, -5.0, 1.5,2.5)\\\nonumber
\end{eqnarray}
\vspace{-1.7cm}
\begin{eqnarray}\nonumber
{\boldsymbol{\theta}}_{(J=4)}&=&(\alpha_{2}, \alpha_{3}, \alpha_{4}, \beta_{12}, \beta_{13}, \beta_{14}, \beta_{22}, \beta_{23}, \beta_{24})\\\nonumber
&=&(1.5, 3.0, 2.0, -3.0, -5.0, -4.0, 1.5, 2.5, 3.0)\\\nonumber
\end{eqnarray}
\vspace{-1.7cm}
\begin{eqnarray}\nonumber
{\boldsymbol{\theta}}_{(J=5)}&=&(\alpha_{2}, \alpha_{3}, \alpha_{3}, \alpha_{4}, \beta_{12}, \beta_{13}, \beta_{14},\beta_{15}, \beta_{22}, \beta_{23}, \beta_{24},\beta_{25})\\\nonumber
&=&(1.5, 3.0, 2.0, 4.0, -3.0, -5.0, -4.0, 1.5, 2.5,3.0, 3.5)\nonumber
\end{eqnarray}

All simulations were implemented in \texttt{R} software \cite{team2020}, using the \texttt{nnet} package to fit the multinomial models \cite{ripley2016package}, and the \texttt{hnp} package \cite{moral2017half} (\texttt{hnp} function) to generate the half-normal plots with a simulated envelope.

\subsection{Results for individual data}

\begin{figure}[!thb]
\centering
\includegraphics[width=0.8\textwidth]{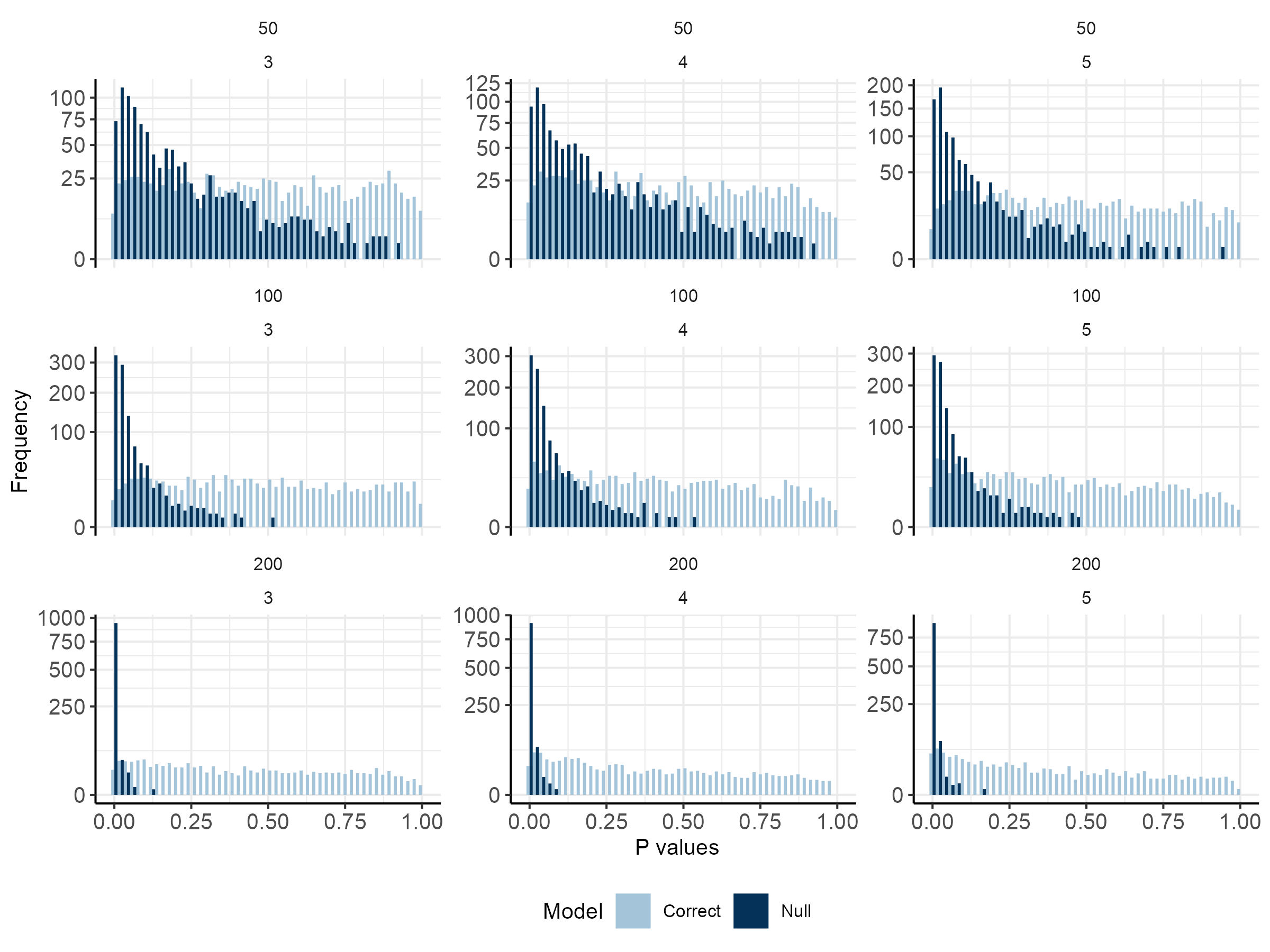}
\caption{Histograms of p-values obtained via the Shapiro-Wilk test for the standardized randomized quantile residuals for $1,000$ simulations when fitting (a) the null model (intercept only), and (b) model 1 (including continuous covariate; correct linear predictor) for $N= 50, 100, 200$ and $J=3,4,5$.}
\label{figRQRcont}
\end{figure}

\begin{figure}[!thb]
\centering
\includegraphics[width=0.8\textwidth]{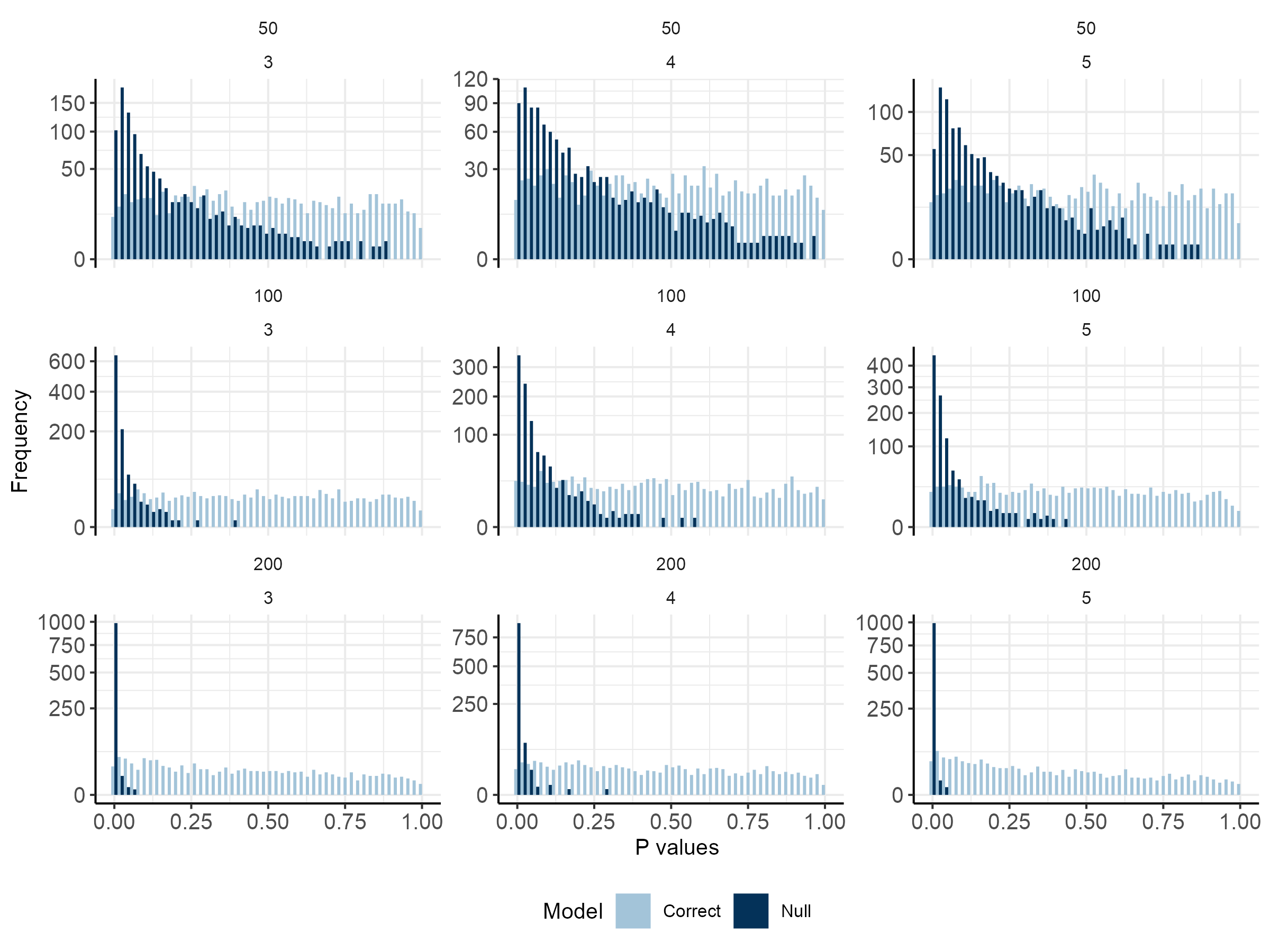}
\caption{Histograms of p-values obtained via the Shapiro-Wilk test for the standardized randomized quantile residuals for $1,000$ simulations when fitting (a) the null model (intercept only), and (b) model 2 (including continuous and dummy covariates; correct linear predictor) for $N= 50, 100, 200$ and $J=3,4,5$.}
\label{figRQRcontcateg}
\end{figure}

We firstly compare residuals obtained from fitting model 1 to the data generated by model 1 itself, and from fitting the null model (intercept only; scenario 1). The distribution of the p-values of the Shapiro-Wilk test are presented in Figure \ref{figRQRcont}, in which we observe that the residuals under the null model are considered to be mostly not normal, while a uniform pattern is seen for the p-values for the residuals obtained from the correct model. Similar patterns are observed for the scenario where model 2 was considered (scenario 2; Figure~\ref{figRQRcontcateg}). It should also be noted, in both scenarios (model 1 and model 2), that the number of categories has no influence on the residual analysis, unlike the influence of the sample size, but this is also related to the sensitivity of the Shapiro-Wilk test. 
Specifically, the normality of residuals was rejected by the Shapiro-Wilk test $p<0.05$) in most simulations considering the null model. As illustration, for example, with $J=3$ and $N=50$, normality was rejected $86.6\%$ of the time when considering model 1, and $92.7\%$ of the time when considering model 2. However, when considering the correct linear predictors, normality was rejected only for $4.0\%$ and $5.7\%$ of the simulated datasets for models 1 and 2, respectively (i.e., close to 5\%, as expected). This shows we may identify lack-of-fit of a multinomial model fitted to individual data by analysing the normality of the randomized quantile residuals.

\newpage
\subsection{Results for grouped data}

\begin{figure}[!thb]
\centering
\includegraphics[width=0.9\textwidth]{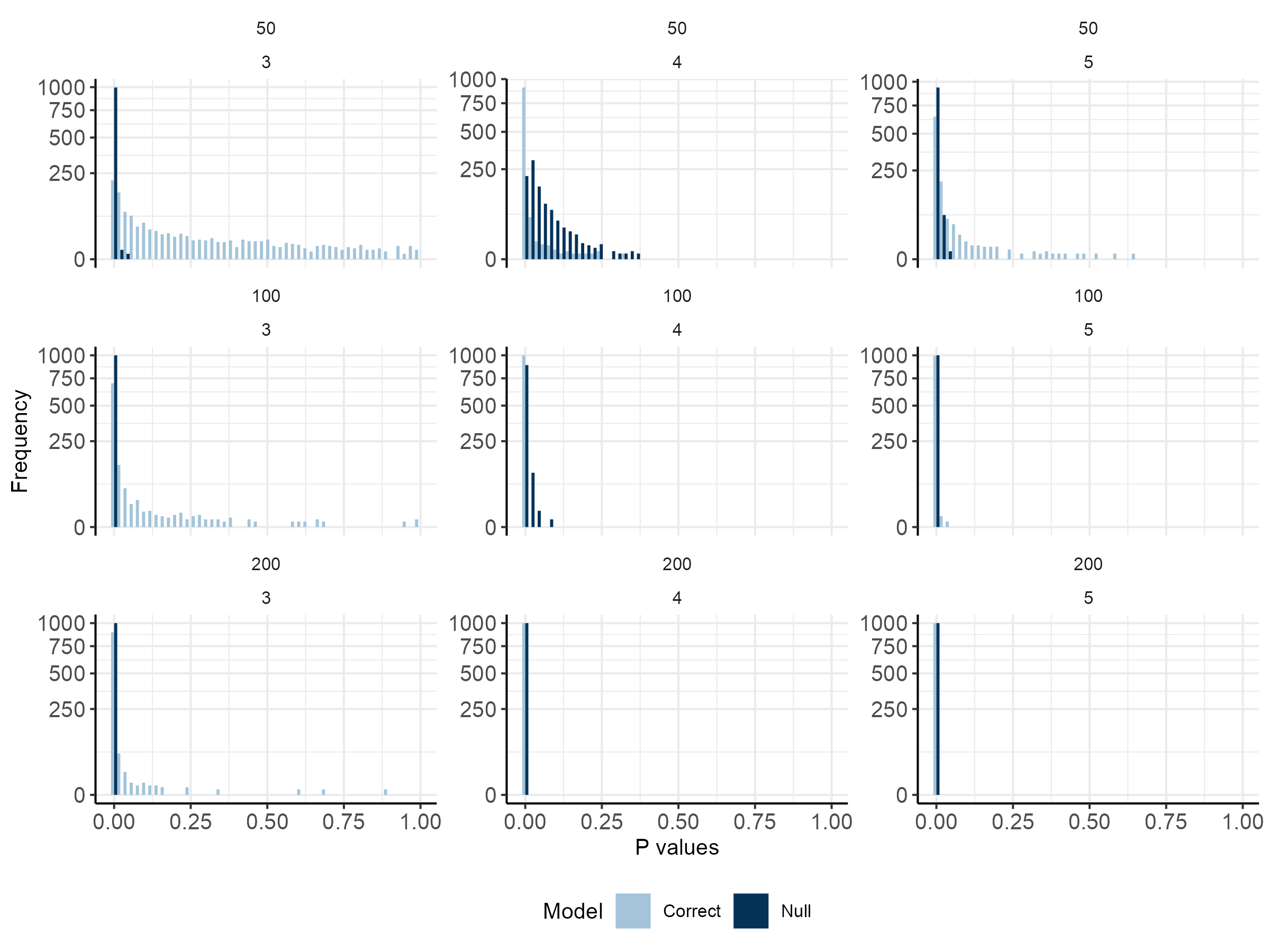}
\caption{Histograms of p-values obtained via the Shapiro-Wilk test for the standardized randomized quantile residuals for $1,000$ simulations when fitting (a) the null model (intercept only), and (b) model 1 (including continuous covariate; correct linear predictor) with grouped data~($m=10$), for $N= 50, 100, 200$ and $J=3,4,5$.}
\label{figRQRcont_group}
\end{figure}

\begin{figure}[!thb]
\centering
\includegraphics[width=0.9\textwidth]{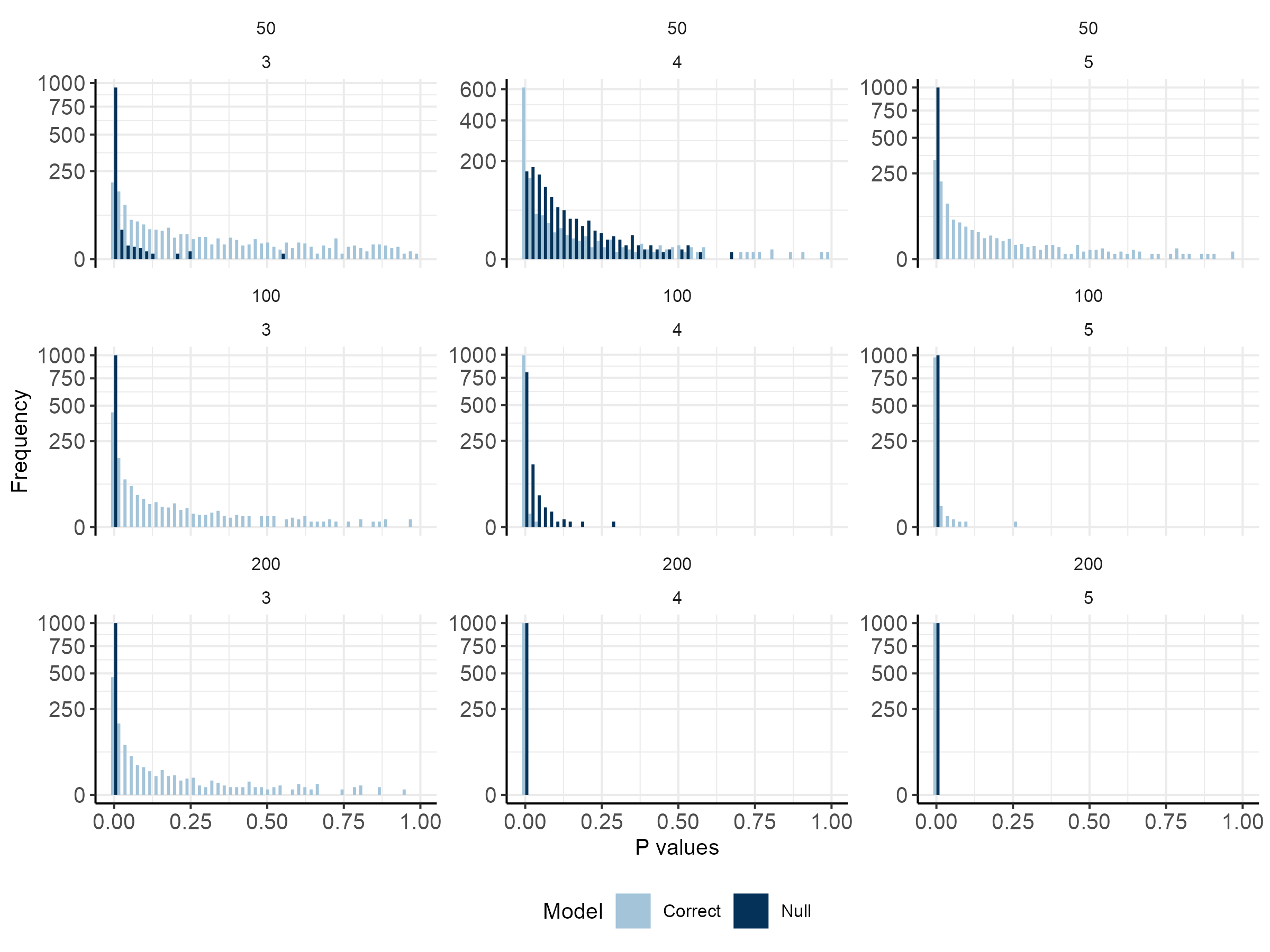}
\caption{Histograms of p-values obtained via the Shapiro-Wilk test for the standardized randomized quantile residuals for $1,000$ simulations when fitting (a) the null model (intercept only), and (b) model 2 (including continuous and dummy covariates; correct linear predictor) with grouped data~($m=10$), for $N= 50, 100, 200$ and $J=3,4,5$.}
\label{figRQRcontcateg_group}
\end{figure}

The results for the grouped data, in general, were similar for all $m$ values, indicating that the group dimension did not represent a source of variation for the standardized randomized quantile residuals and also to the Euclidean and Mahalanobis distance measures, particularly in this study. 
In this way, we present here the results for $m=10$, with the others results available at \url{https://github.com/GabrielRPalma/DiagnosticsForCategoricalResponse}. 
Initially, we present the distribution of p-values referring to the Shapiro-Wilk test applied to the quantile residuals for grouped data, considering scenario 3~(model 1 versus null) and scenario 4~(model 2 versus null). Just as in the individual case the results were satisfactory, i.e,  the normality of residuals was rejected by the Shapiro-Wilk test $p<0.05$) in most simulations considering the null model, as can be observed from Figures \ref{figRQRcont_group}~(scenario 3) and  \ref{figRQRcontcateg_group}~(scenario 4).

\begin{figure}[!thb]
\centering
\includegraphics[width=0.8\textwidth]{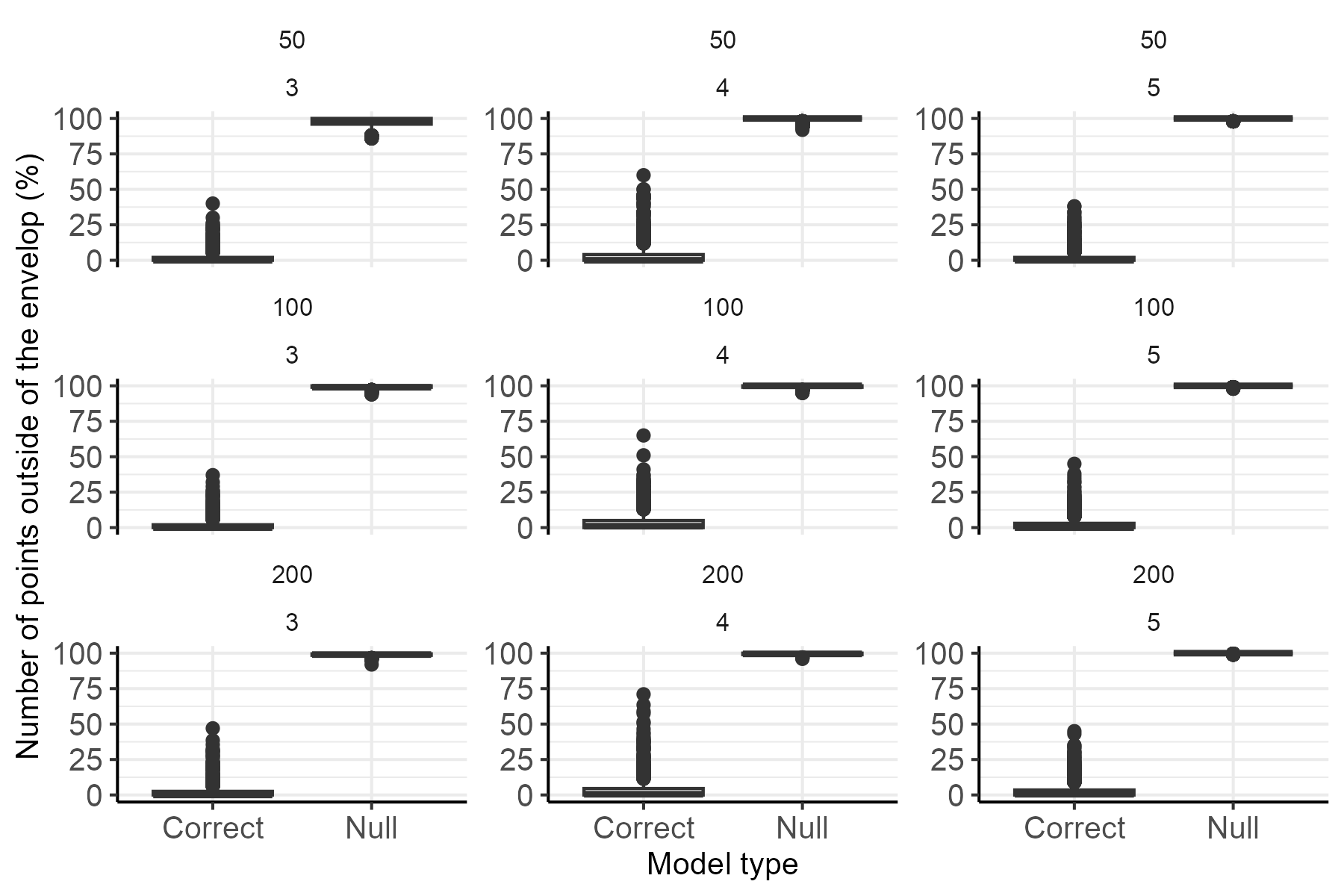}
\caption{Boxplots of the percentage of points outside the simulated envelope for model 1 (including continuous covariate) and the null model (intercept only) using the Euclidean  distance, for $N= 50, 100, 200$ and $J=3,4,5$.}
\label{figDistcontE}
\end{figure}

Next, we present the results for distances measures, considering model 1 versus null and Euclidean distance~(scenario 5) and Mahalanobis distance~(scenario 6). For both scenarios, it was possible to distinguish the true model from the null model by using half-normal plots with a simulated envelope for the distances (Figure~\ref{figDistcontE}).

\begin{figure}[!thb]
\centering
\includegraphics[width=0.85\textwidth]{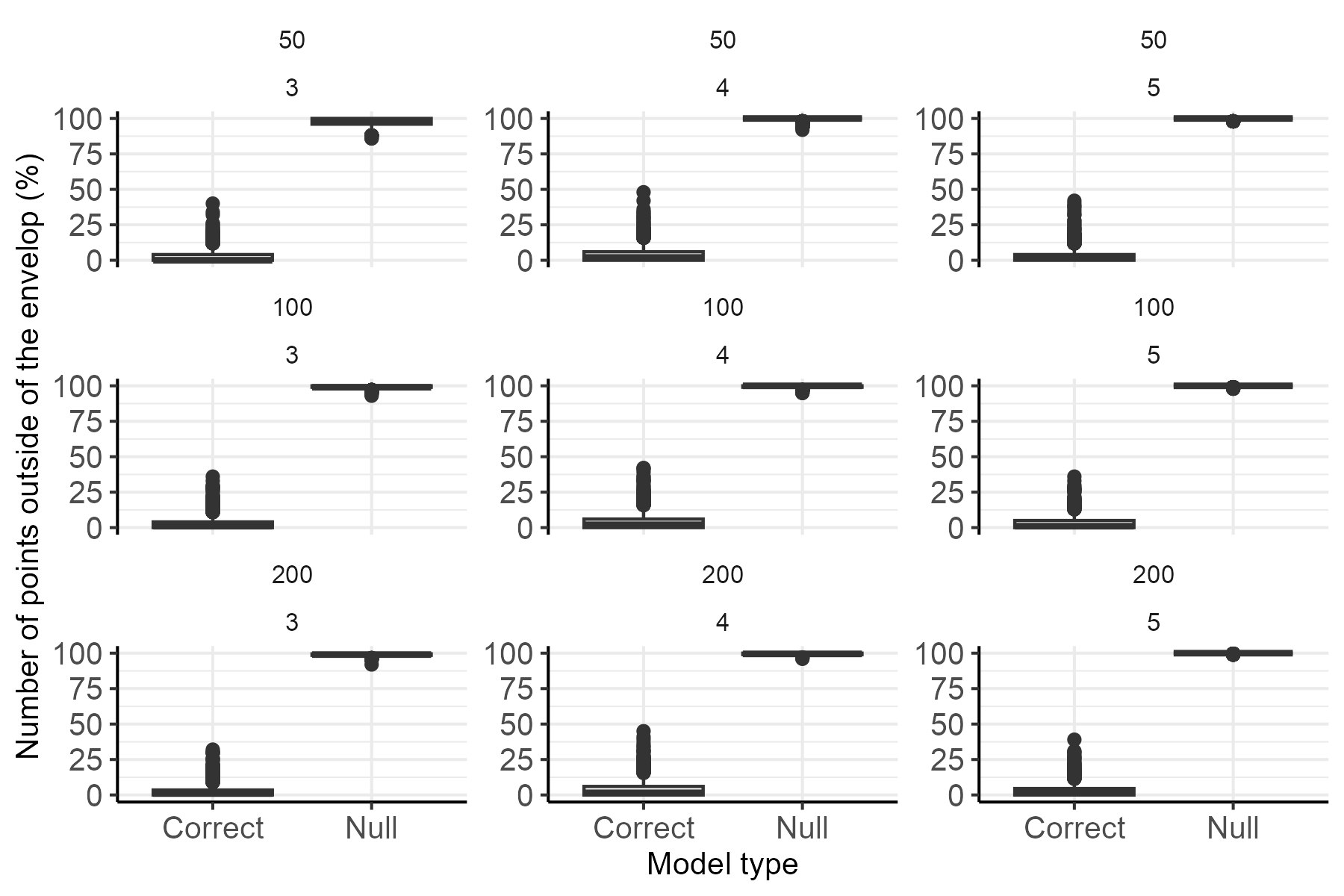}
\caption{Boxplots of the percentage of points outside the simulated envelope for model 1 (including continuous covariate) and the null model (intercept only) using the Mahalanobis  distance, for $N= 50, 100, 200$ and $J=3,4,5$.}
\label{figDistcontM}
\end{figure}

\begin{figure}[!thb]
\centering
\includegraphics[width=0.8\textwidth]{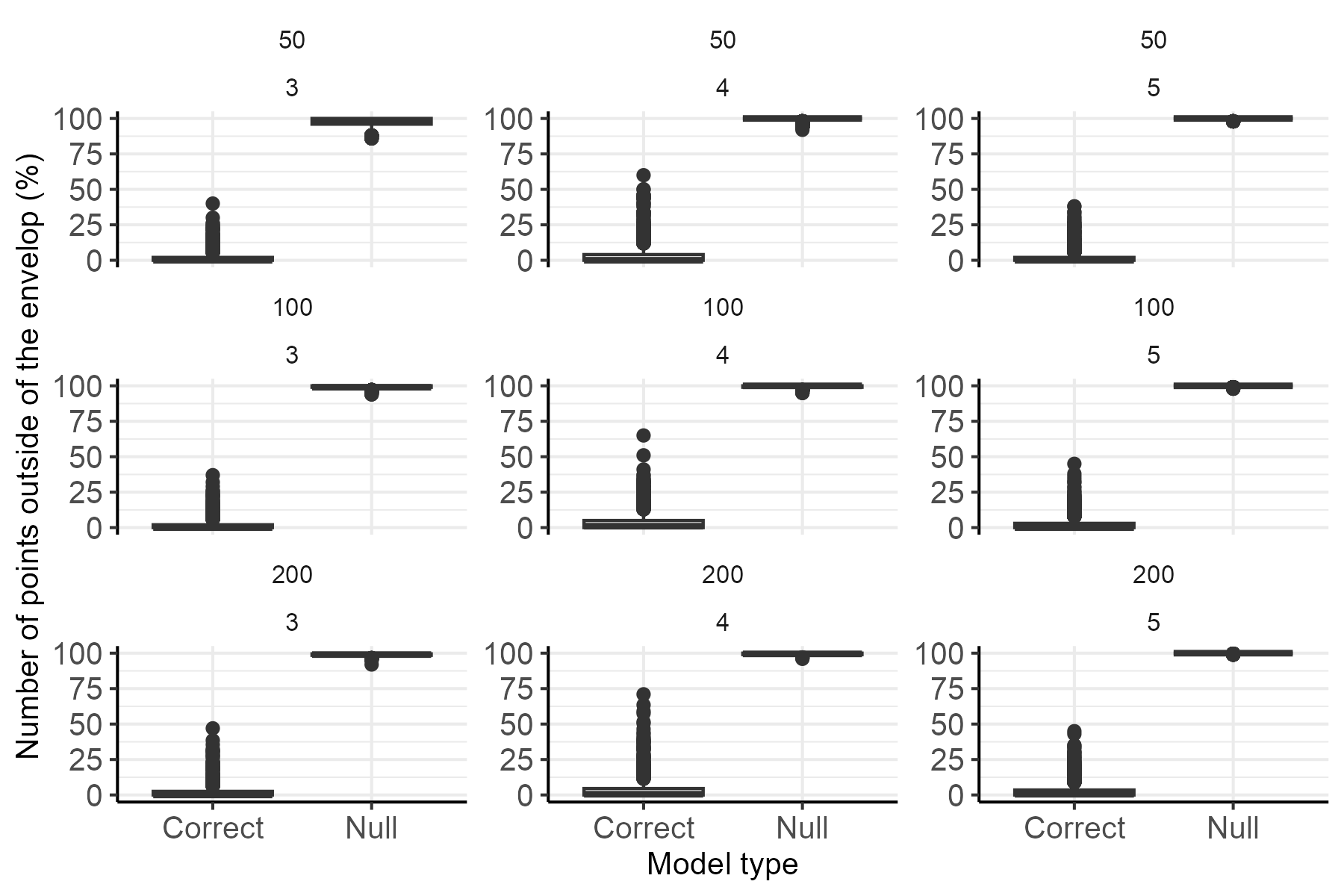}
\caption{Boxplots of the percentage of points outside the simulated envelope for model 2 (including continuous and dummy covariates) and the null model (intercept only) using the Euclidean  distance, for $N= 50, 100, 200$ and $J=3,4,5$.}
\label{figDistcontEm2}
\end{figure}

\begin{figure}[!thb]
\centering
\includegraphics[width=0.85\textwidth]{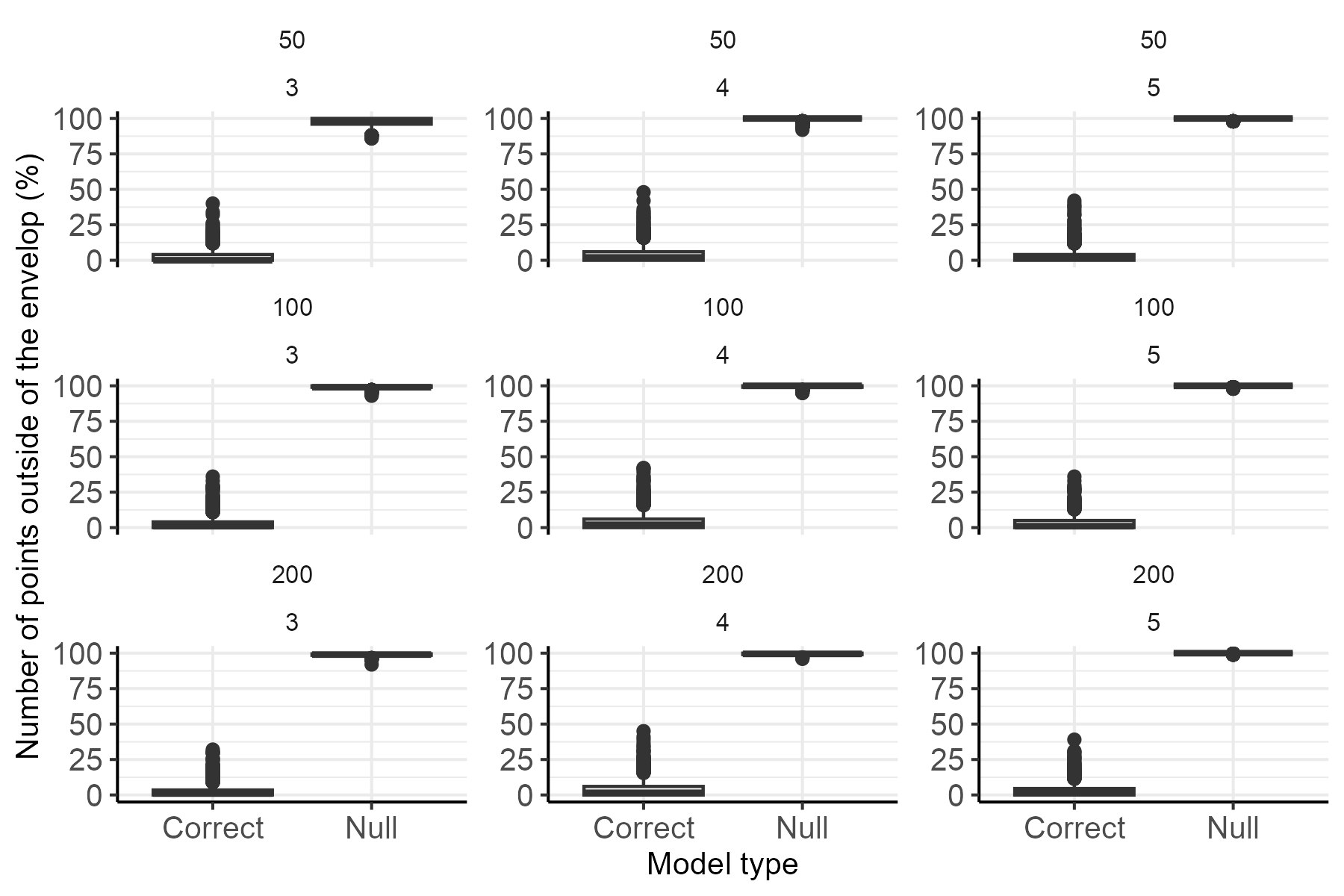}
\caption{Boxplots of the percentage of points outside the simulated envelope for model 2 (including continuous and dummy covariates) and the null model (intercept only) using the Mahalanobis  distance, for $N= 50, 100, 200$ and $J=3,4,5$.}
\label{figDistcontMm2}
\end{figure}

The median of the percentage of points outside the envelope is less than 5\% for model 1, considering both distances, as opposed to almost 100\% for the null model. Also, the distribution of these values within each level appears to be symmetric and has approximately the same variability (Figures \ref{figDistcontE} and \ref{figDistcontM}). Similar conclusions can be drawn for model 2 for the Euclidean Figure~\ref{figDistcontEm2}) and Mahalanobis distances (Figure~\ref{figDistcontMm2}), given that the median of the percentage of points outside the envelope is less than 5\% for model 2 and close to 100\% for the null model using both distances.

This confirms that the proposed diagnostics are useful to identify well-fitting multinomial models for grouped nominal data.

\section{Applications} \label{aplic}

Here, two motivation studies  available in the literature are considered to illustrate the procedures presented in Sections \ref{qr} and \ref{dist}.

\subsection{Wine Classification}

This first dataset (individual structure) arises from a study carried out by \cite{forina1991parvus}, involving wine classification techniques (\cite{jing2010graphical,ahammed2018predicting}). In this study, a chemical analysis was carried out at the Institute of Pharmaceutical and Food Analysis and Technologies about 178 wines from three grape cultivars from the Liguria region in Italy, whose objective was to classify the different cultivars. The response variable represents the type of cultivar, assuming values $\left\{{1,2,3}\right\}$. In the analysis, the amounts of 13 chemical constituents of each cultivar were determined, among which are magnesium and phenols that can be considered good indicators of wine origin \cite{kallithraka2001instrumental}. Further details as well as the dataset are available in the \texttt{rattle.data} \cite{rattle.data2011} package for \texttt{R} software \cite{team2020}.

We define the following linear predictors:
$M1$: intercept only (null model); $M2$: intercept + phenols;
$M3$: intercept + magnesium + phenols (additive model) and
$M4$: intercept + magnesium * phenols (interaction model).

The final model was selected by applying likelihood-ratio (LR) tests to a sequence of nested models and we obtained: $M1 \times M2$ LR $=123.98~(p<0.01)$;
$M2 \times M3$ LR $=13.14~(p<0.01)$ and 
$M3 \times M4$ LR $=1.25~(p=0.54)$,
all statistics associated to 2 degrees of freedom. Therefore, model M3 was selected. The Akaike Information Criterion (AIC) also was used to compare models, and the lowest AIC (261.50) value was for model M3, but this measure does not verify the goodness-of-fit of the model or validates the distributional assumption. 


\begin{figure}[!thb]
\centering
\includegraphics[width=0.8\textwidth]{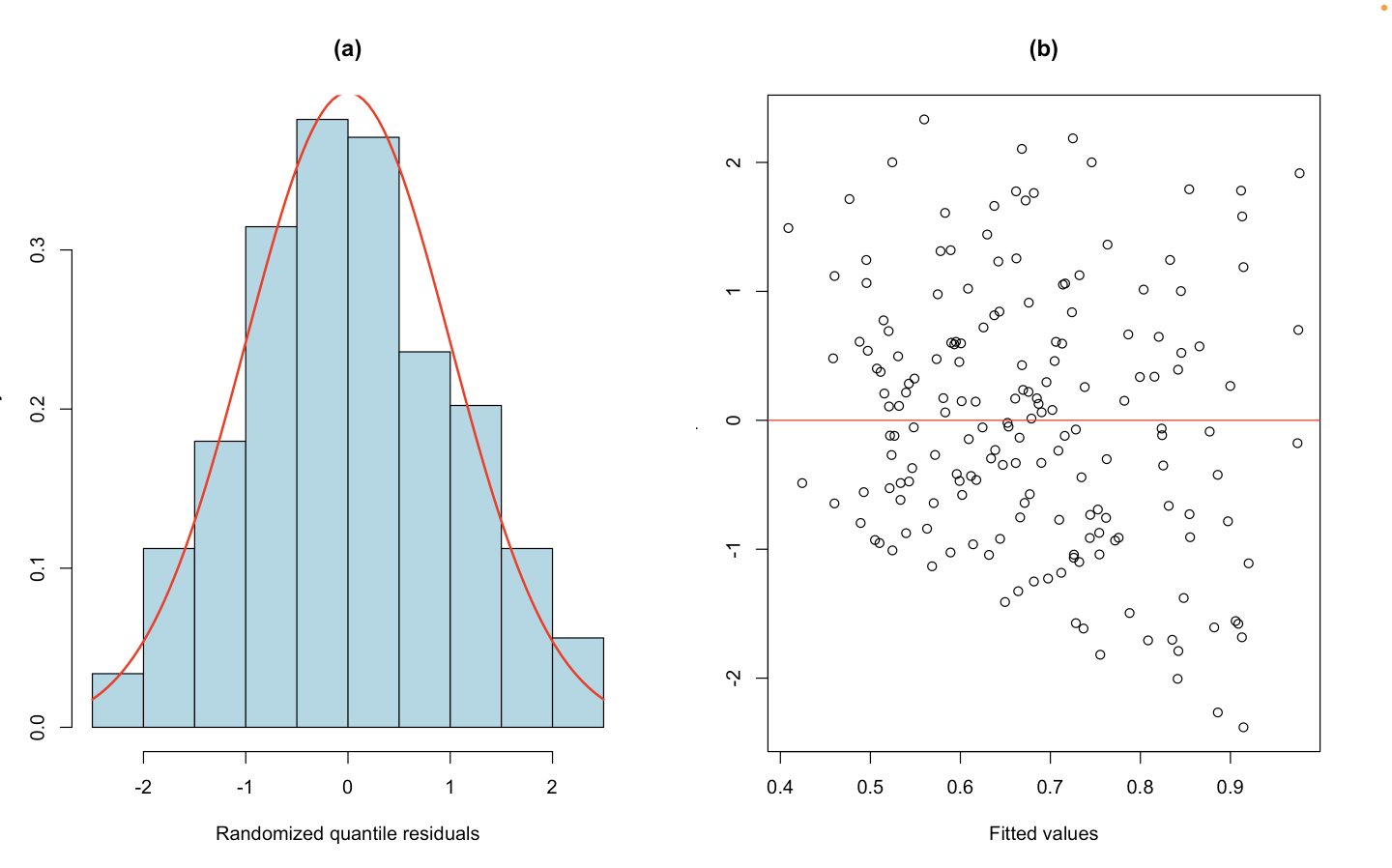}
\caption{Distribution of standardized randomized quantile residuals from model M3~(a standard normal density curve in red) and   fitted values versus  standardized randomized quantile residuals, according to the wine study data (\cite{forina1991parvus}).}
\label{Hist_vinho}
\end{figure}

The histogram of the randomized quantile residuals (Figure \ref{Hist_vinho}(a)) indicate that resdiduals of model M3 are normally distributed. This is confirmed by the Shapiro-Wilk test ($p=0.167$). Also, the plot of residuals versus fitted values (Figure \ref{Hist_vinho}(b)) Residuals vary mainly between $-2$ and $2$ and no pattern is evident, which also suggests that model M3 is well-fitted to the data.

\begin{figure}[!thb]
\centering
\includegraphics[width=1\textwidth]{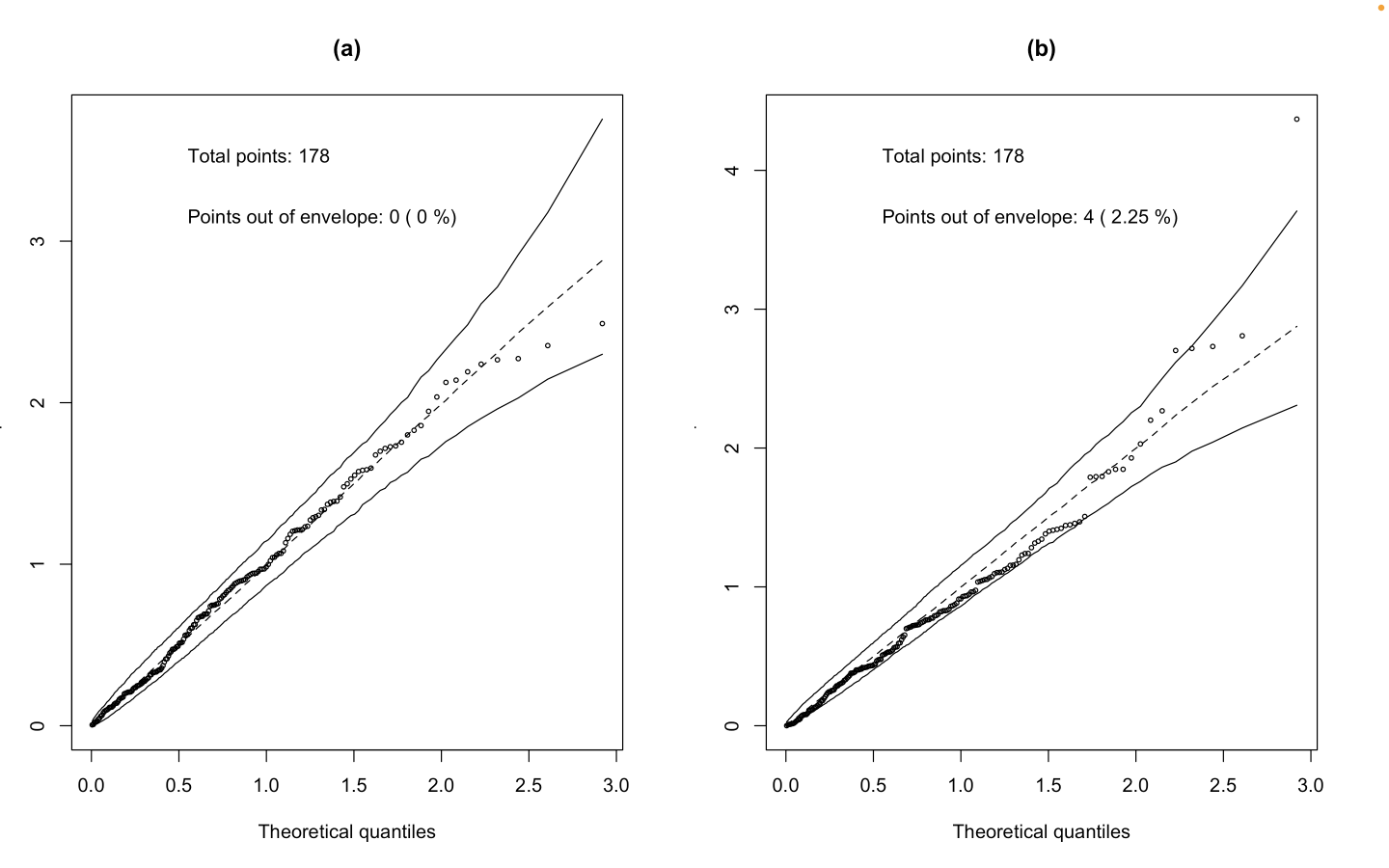}
\caption{ Half-normal plot with a simulated envelope (confidence level = 95\%) of the standardized randomized quantile residuals for model M3:intercept + magnesium + phenols~(a) and the null model~(b) applied to the wine study data \cite{forina1991parvus}.}
\label{fig13}
\end{figure}

The half-normal plots with a simulated envelope for the standardized randomized quantile residuals are shown in Figure \ref{fig13} for model M3:~ intercept + magnesium + phenols~(a) and the null model~(b). It appears that the model fits the data well, since no point is outside the envelope.  

\subsection{Student preference}

The second dataset (grouped structure) refers to the choice made by high school students among different programs. This sample of 200 individuals was made available in 2013 by the statistical consulting group at the University of California at Los Angeles (UCLA), being used in studies involving polytomous data (e.g. \cite{molina2015multinomial}, \cite{abonazel2018liu} and \cite{dalzell2018regression}). The response variable is the choice by a program~($1$: academic, $2$: general, $3$: vocational). There are $11$ covariates available in this study, including socioeconomic status, gender, and scores in specific subjects (mathematics, social studies, writing, among others). Here, we consider the maths score as a continuous covariate, to verify if the score contributed to the student’s decision. The data were organized in $N=34$ including groups varying from $m=2$ to $m=13$. For more details, see \cite{hsbdemo2021}.



Considering the null hypothesis that program choice is independent of maths score, we employed a LR test to compare a null model (intercept only) with a model including the maths score in the linear predictor (model M1). We obtained a test statistic of $51.97$, on 2 degrees of freedom ($p<0.01$). Model M1 also presented a lower AIC (182.81) when compared to the null model (230.77). Based on this result, it is concluded that maths score is significant to explain program choice. 

\begin{figure}[!thb]
\centering
\includegraphics[width=1\textwidth]{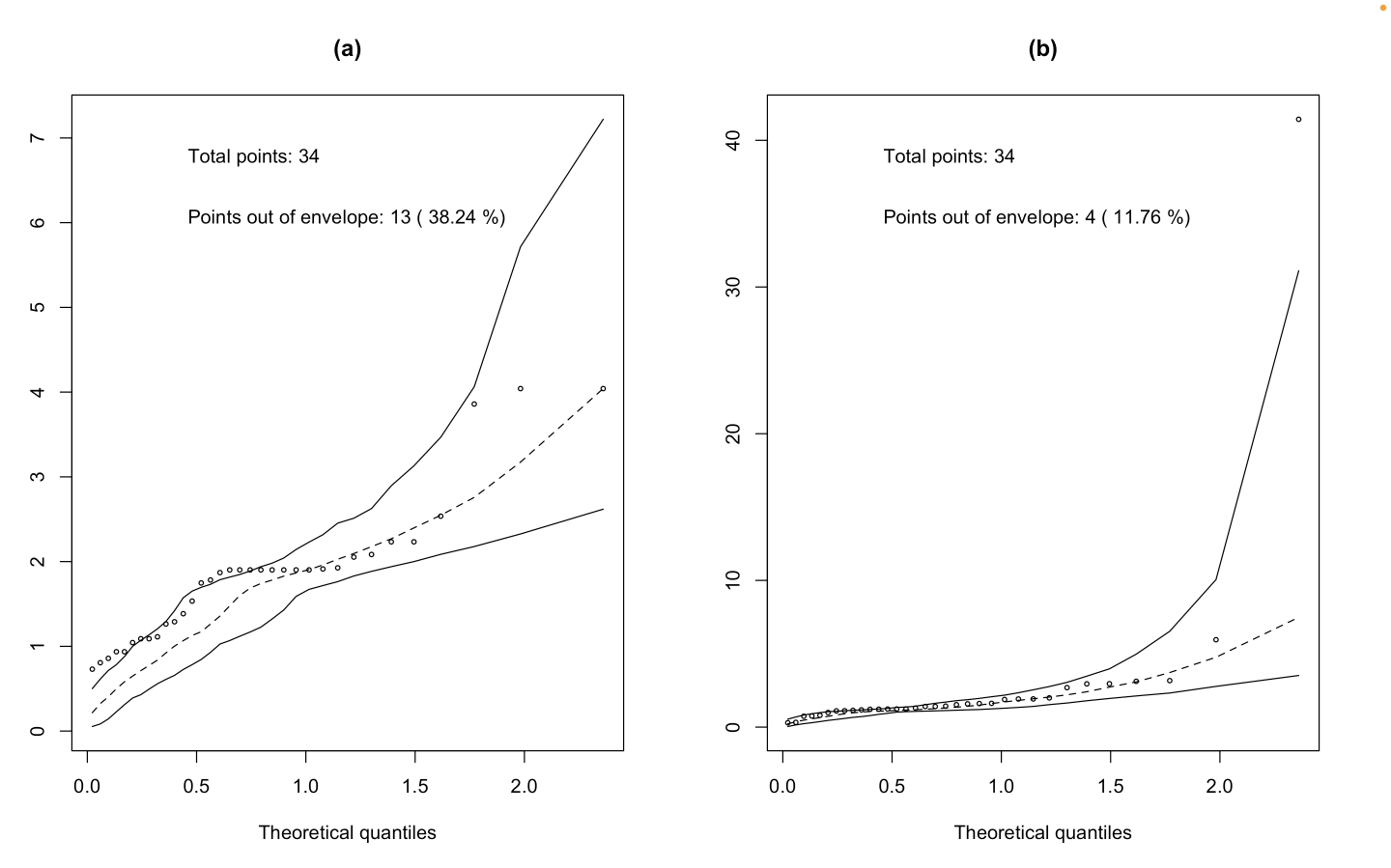}
\caption{ Half-normal plot with a simulated envelope (confidence level = 95\%) using Euclidean distance for null model~(a) and M1: intercept + maths score~(b) for the data available in \cite{hsbdemo2021}.}
\label{fig_Dist_half_Eucl}
\end{figure}
\begin{figure}[!thb]
\centering
\includegraphics[width=1\textwidth]{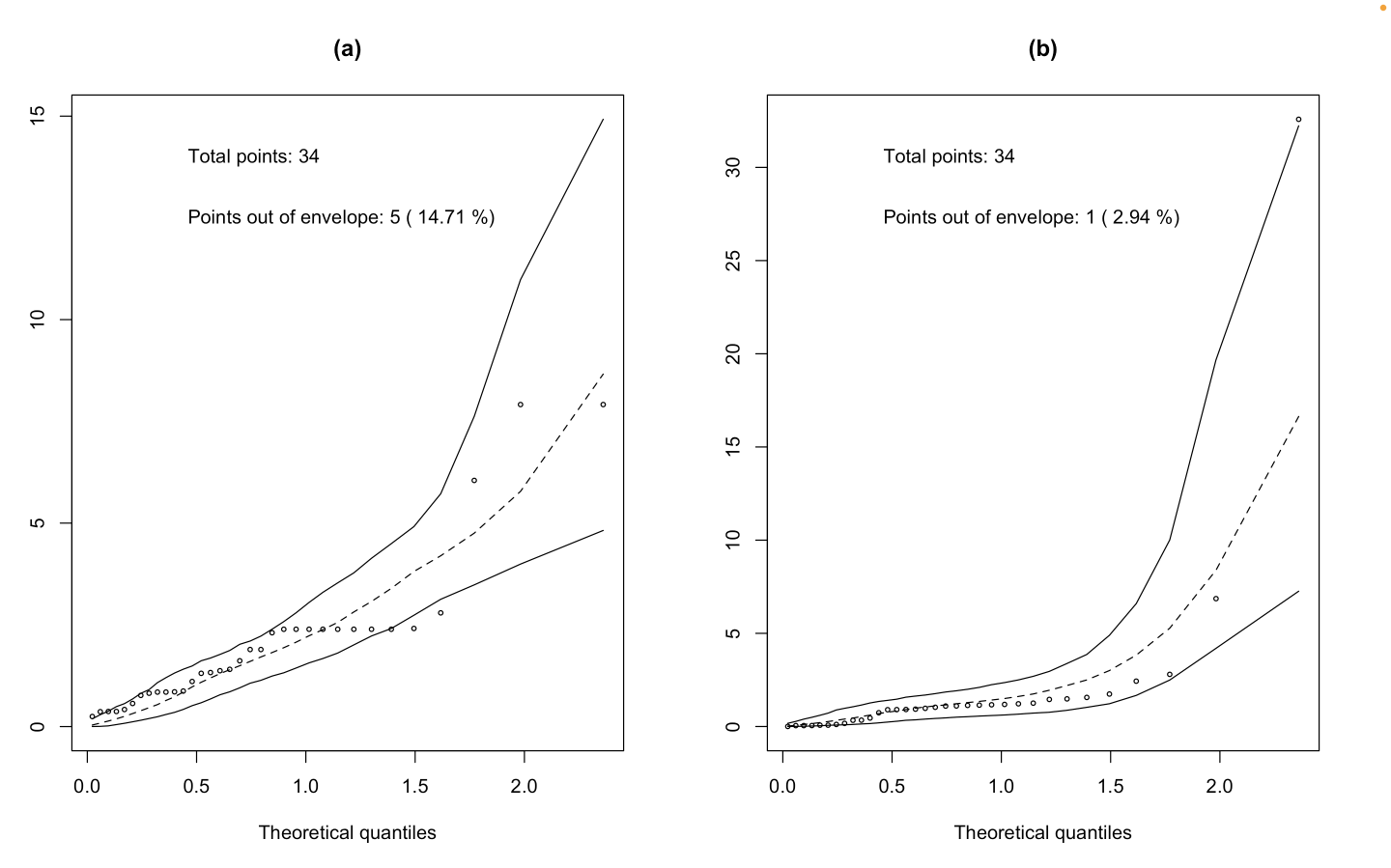}
\caption{ Half-normal plot with a simulated envelope (confidence level = 95\%) using Mahalanobis distance for null model~(a) and M1: intercept + maths score~(b) for the data available in \cite{hsbdemo2021}}
\label{fig_Dist_half_Mah}
\end{figure}

The half-normal plots with a simulated envelope indicate that model M1~(intercept + maths score) is suitable to analyse the data, for both Euclidean~(Figure \ref{fig_Dist_half_Eucl} and Mahalanobis~(Figure \ref{fig_Dist_half_Mah}) distances.

\section{Conclusion} \label{final}

In this work we presented alternatives to residual analysis for nominal data with individual and grouped data structures using randomized quantile residuals and distance measures, respectively. The simulation studies showed that these residuals and the proposed distances presented good performance in assessing model goodness-of-fit with continuous and categorical covariates. Therefore, the randomized quantile residuals and the distances may be potential tools for checking diagnostics of generalized logit models. However, the analysis of residuals for polytomous data has many challenges yet to be explored. Studies focusing on small sample sizes are necessary to assess the fit of the model, which could lead to sampling uncertainty in the residuals and  distances. Venues for future work also include simulation studies focusing on longitudinal designs.

\section*{Acknowledgments}
This work derived from the thesis entitled ``Residuals and diagnostic methods in models for polytomous data'' with support from the Brazilian Foundation, Coordenação de ``Coordena\c{c}\~{a}o de Aperfei\c{c}oamento de Pessoal de N\'{i}vel Superior'' (CAPES)  process number $88882.378344/2019-01$.
This publication also had the additional support from
Brazilian Fundation-CAPES  process number $88887.716582/2022-00$ and 
 from  Science Foundation Ireland under grant number $18/CRT/6049$.  

\section*{Supplementary material}
All R code, including the implementations of the proposed methods, are available at \url{https://github.com/GabrielRPalma/DiagnosticsForCategoricalResponse}.

\bibliographystyle{tfs}
\bibliography{interacttfssample}

\end{document}